\documentclass{jfm}
\usepackage{natbib}
\usepackage{epsfig}
\usepackage{amssymb}

\def\begineq{\begin{equation}}
\def\endeq{\end{equation}}
\def\be{\begin{equation}}
\def\ee{\end{equation}}

\title[Energy dissipation in body-forced plane shear flow]
{Energy dissipation in body-forced plane shear flow}
\author[C.R. Doering, B. Eckhardt and J. Schumacher]%
{C.\ns R.\ns D\ls O\ls E\ls R\ls I\ls N\ls G$^{1,2}$,\ns
 B.\ns E\ls C\ls K\ls H\ls A\ls R\ls D\ls T$^3$ \and
 J.\ns S\ls C\ls H\ls U\ls M\ls A\ls C\ls H\ls E\ls R$^3$}
\affiliation{$^1$ Department of Mathematics, University of Michigan,
Ann Arbor, MI 48109-1109,  USA\\[\affilskip]
$^2$ Michigan Center for Theoretical Physics,
Ann Arbor, MI 48109-1120,  USA\\[\affilskip]
$^3$ Fachbereich Physik, Philipps-Universit\"at, D-35032 Marburg, Germany}
\pubyear{2003}
\volume{999}
\pagerange{1--10}
\date{May 14, 2003}
\begin{document}
\maketitle
\begin{abstract}
We study the problem of body-force driven shear flows in a plane channel 
of width $\ell$ with free-slip boundaries.  
A mini-max variational problem for upper bounds on the bulk time averaged 
energy dissipation rate $\epsilon$ is derived from the incompressible 
Navier-Stokes equations with no secondary assumptions.
This produces rigorous limits on the power consumption that are valid for 
laminar or turbulent solutions.
The mini-max problem is solved {\it exactly} at high Reynolds numbers $Re 
= U\ell/\nu$, where $U$ is the rms velocity and $\nu$ is the kinematic 
viscosity, yielding an explicit bound on the dimensionless asymptotic 
dissipation factor $\beta=\epsilon \ell/U^3$ that depends only on the 
``shape'' of the shearing body force.
For a simple half-cosine force profile, for example, the high Reynolds 
number bound is $\beta \le \pi^2/\sqrt{216} = .6715\dots$.
We also report extensive direct numerical simulations for this particular 
force shape up to $Re \approx 400$; the observed dissipation rates are 
about a factor of three below the rigorous high-$Re$ bound.
Interestingly, the high-$Re$ optimal solution of the variational problem 
bears some qualitative resemblence to the observed mean flow profiles in 
the simulations. 
These results extend and refine the recent analysis for body-forced 
turbulence in J. Fluid Mech. {\bf 467}, 289-306 (2002).
\end{abstract}

\section{Introduction}
Bounds on the energy dissipation rate for statistically stationary flows 
belong to the small class of rigorous results for turbulence that can be 
derived directly from the incompressible Navier-Stokes equations without 
introducing any supplementary hypotheses or uncontrolled approximations. 
Quantitative approaches are mostly based on variational formulations as 
have been used in a variety of boundary-driven turbulent flows; see, e.g., 
\cite{Howard72}, \cite{Busse78}, \cite{Doering94}, \cite{Nicodemus98}, 
\cite{Kerswell98}.
More recently \cite{Childress2001} and \cite{Doering02} extended these 
analyses to body-forced flows in a fully periodic domain.
The motivation for such studies is to consider mathematically well-defined 
and tractable models for (almost) homogeneous and (almost) locally 
isotropic stationary turbulence when boundaries are far away.

Define the Reynolds number $Re=U\ell/\nu$, where $U$ is the steady state 
rms velocity, $\ell$ is the longest characteristic length scale in the 
body-force function and $\nu$ is the kinematic viscosity. 
For the three-dimensional Navier-Stokes equations, \cite{Doering02} found 
that the energy dissipation rate per unit mass $\epsilon$ satisfies
\begin{equation}
\epsilon \le c_1\nu \frac{U^2}{\ell^2}+c_2 \frac{U^3}{\ell}
\label{upperb} 
\end{equation} 
where the coefficients $c_1$ and $c_2$ depend only on the functional shape 
of the body-force, and not on any other parameters or on any ratios 
involving the (say, rms) amplitude  $F$ of the force or the overall system 
size $L$---which could be arbitrarily larger than $\ell$.
(We give a more precise definition of the ``shape'' of the forcing 
function below, or else see \cite{Doering02}.)
In terms of the dimensionless dissipation ratio $\beta = \epsilon 
\ell/U^3$, this result is $\beta \le c_1/Re+c_2$, an estimate in 
qualitative accord with theoretical, computational, and experimental 
result for homogeneous isoptropic turbulence (\cite{Frisch95}, 
\cite{Sreeni84}, \cite{Sreeni98}).
The analysis in \cite{Childress2001} focuses on dissipation estimates in 
terms of the true control parameter for such systems, the Grashof number 
$Gr=F \ell^3/\nu^2$, but those results are less easily interpreted in 
terms of conventional ideas for homogeneous isotropic turbulence.

In this paper we refine and develop the approach in \cite{Doering02} for 
the particular example of flow between free-slip boundary planes driven by 
a steady volume forcing density.
Homogeneous shear turbulence is of interest in its own right, and by 
setting a uniform direction of the force we simplify some of the analysis 
allowing us to improve the bounds and make quantitative comparison with 
direct numerical simulations.
The analysis produces rigorous limits that are approached within about a 
factor of three even at the moderate values of $Re$ (up to 400) that we 
are able to reach computationally (in these runs the usual Taylor 
microscale Reynolds number $R_{\lambda} \approx 100$).

The rest of this paper is outlined as follows.
In the next section we describe the model in detail and introduce some 
notation and definitions. 
In section 3 we derive the mini-max problem for the dissipation rate 
bounds.
Elementary analysis quickly produces estimates which are then refined to 
the optimal bound (within this variational formulation) in the limit of 
$Re\to\infty$. 
The final section 4 is a comparison of the results with the computational 
data and a brief discussion of the results.

\section{Preliminaries}
The three-dimensional dynamics of the flow is governed by the 
Navier-Stokes equations for an incompressible Newtonian fluid,  
\begin{eqnarray}
\label{nseq}
\frac{\partial{\bf u}}{\partial t}+({\bf u}\cdot{\bf \nabla}){\bf u}
+ {\bf \nabla} p = \nu {\bf \nabla}^2{\bf u}+{\bf f}, \ \ \ \ \ 
{\bf \nabla}\cdot{\bf u}&=&0,
\end{eqnarray}
where the velocity field is ${\bf u}({\bf x},t) = {\bf e}_x u_x +
{\bf e}_y u_y + {\bf e}_z u_z$, the
pressure field is $p({\bf x},t)$, and the kinematic viscosity is $\nu$.
The flow is in the slab, $[0,L_x]\times [0,\ell]\times [0,L_z]$  
with periodic boundary conditions in the $x$ (streamwise) 
and $z$ (spanwise) directions. 
On the bottom and top surfaces at $y=0$ and $y=\ell$, we take the 
free-slip boundary conditions $u_y=0$ and $\partial_y u_x = 0 = \partial_y 
u_z$.

The steady body-force shearing the fluid is taken to be of the form
\begin{eqnarray}
\label{force1}
{\bf f}({\bf x})=
F \phi\left(\frac{y}{\ell}\right){\bf e}_x\,.
\end{eqnarray}
The length scale $\ell$ is the longest length scale in the forcing function.
The dimensionless {\it shape} function $\phi: [0,1] \rightarrow {\cal R}$ 
satisfies homogeneous Neumann boundary conditions with zero mean:
\begin{equation}
\phi'(0) = 0 =\phi'(1), \ \ \ \ \ \ \int_0^1{\phi(\eta) d\eta} = 0.
\label{mz}
\end{equation} 
Technically we require that $\phi$ is a square integrable function, i.e., 
$\phi \in L^2[0,1]$, but in practice we are interested in even smoother 
functions whose Fourier transforms are effectively supported on low 
wavenumbers.
The amplitude $F$ is specified uniquely for a given ${\bf f}$ when we fix 
the normalization of the shape function by
\begin{equation}
1 = \int_0^1{\phi(\eta)^2 d\eta}\,.
\end{equation}
We also introduce the dimensionless ``potential'' for the body-force shape 
function via
\begin{equation}
{\bf f}({\bf x}) = \nabla \times \left[- F \Phi\left(\frac{y}{\ell}\right) 
{\bf e}_z \right]
\,.
\end{equation}
Then $\Phi \in H^1[0,1]$, the space of functions with square integrable 
first derivatives, so $\phi = -\Phi^{\prime}$ and, without loss of 
generality because $\phi$ has zero mean, it satisfies homogeneous 
Dirichlet boundary conditions, $\Phi(0) = 0 = \Phi(1)$.

At sufficiently high forcing amplitude, a finite perturbation causes 
transition to turbulence and the imposed driving sustains the turbulent 
state assuring statistical stationarity of the turbulent flow. 
In the following $\langle \cdot \rangle$ denotes the space-time average.
Using the root mean square value $U=\sqrt{\langle {\bf u}^2\rangle}$ of 
the total velocity field---including both a possible mean flow and 
turbulent fluctuations---and the length scale $\ell$ in the force, the 
Reynolds number is
\begin{eqnarray}
Re=\frac{U \ell}{\nu} \ .
\end{eqnarray}
The energy dissipation per unit mass is $\epsilon=\nu\langle|\nabla{\bf 
u}|^2\rangle$ and we define the dimensionless dissipation factor $\beta$ 
via \begin{equation}
\epsilon =\beta \ \frac{U^3}{\ell}\,.
\end{equation}
Our aim is to derive bounds on $\beta$ as a function of $Re$ and as a 
functional of the shape $\phi$ of the driving force.

\section{Bounds for the energy dissipation}
The calculation of upper bounds on $\beta$ proceeds in two steps.
First is the derivation of a variational expression and second
is the determination of rigorous estimates for it. 

\subsection{The variational problem}
From the averaged power balance in the Navier-Stokes equations, the energy
dissipation rate per unit mass $\epsilon$ is\footnote{Strictly speaking we
are also assuming that the long time averages exist and that this relation
is an equality for the solutions, rather than just an inequality.  That 
is, for weak solutions of the Navier-Stokes equations it is only known
that $\epsilon \le F \left\langle \phi u_x \right\rangle$.
These mathematical technicalities do not alter the ultimate bounds that we 
will derive in this paper.}  
\begin{equation}
\epsilon = \nu \left\langle\left| \nabla {\bf u} \right|^2 \right\rangle
= F \left\langle \phi u_x \right\rangle.
\label{epsdef}
\end{equation}
Another expression for the forcing amplitude $F$ can be obtained by 
projecting onto the momentum equation.
Specifically, we project the streamwise component of the Navier-Stokes 
equations onto a mean zero multiplier function $\psi \in H^2[0,1]$ (a 
function whose second derivative is square integrable) satisfying 
homogeneous Neumann boundary conditions $\psi'(0) = 0 = \psi'(1)$.
The multiplier function $\psi$ must {\it not} be orthogonal to the shape
function $\phi$; we consider only $\left<\phi \psi \right> \ne 0$.
It is also convenient to introduce the derivative of the multiplier
function, $\Psi = \psi' \in H^1[0,1]$, satisfying homogeneous Dirichlet 
boundary conditions $\Psi(0) = 0 = \Psi(1)$.
The inner product of $\phi$ and $\psi$ is the inner product of $\Phi$ and
$\Psi$, i.e., $\left<\Phi \Psi \right> = \left<\phi \psi \right> \ne 0$.

Take the inner product of the Navier-Stokes equation (\ref{nseq}) with 
$\psi(y/\ell){\bf e}_x$, integrate over the volume utilizing appropriate 
integrations by parts, and take the long time average to obtain the 
relation
\begin{equation}
- \left< \frac{1}{\ell} \psi' u_x u_y \right> 
= \left< \frac{\nu}{\ell^2} \psi'' u_x \right> 
+ F \left< \phi \psi \right>.
\end{equation} 
This may be solved for the strength of the applied force $F$
which when inserted into (\ref{epsdef}) yields
\begin{equation}
\epsilon = 
- \frac{\left<\phi u_x\right> \left<\frac{1}{\ell} \psi' u_x u_y
+ \frac{\nu}{\ell^2} \psi'' u_x \right>}
{\left<\phi \psi\right>}.
\label{epsilon_bound}
\end{equation}  

While the force amplitude $F$ is not explicitly displayed in 
(\ref{epsilon_bound}) anymore, it is implicitely present through the 
constraint that the root mean square value of the velocity field is $U$.
Dividing by $U^3/\ell$, we produce an expression for the dimensionless 
dissipation factor $\beta$,
\begin{equation}
\beta = \frac{\epsilon \ell}{U^3} =
- \frac{\left<\phi (\frac{u_x}{U})\right> \left<\psi' (\frac{u_x}{U})
(\frac{u_y}{U}) + \frac{1}{Re} \psi'' (\frac{u_x}{U}) \right>}
{\left<\phi \psi\right>}.
\label{beta1}
\end{equation}
Changing now to normalized velocities
$u {\bf e}_x+ v {\bf e}_y + w {\bf e}_z
= U^{-1}(u_x {\bf e}_x + u_y {\bf e}_y + u_z {\bf e}_z)$, so that
$\left<u^2+v^2+w^2\right>=1$, and dimensionless spatial coordinates
$\ell^{-1}(x {\bf e}_x + y {\bf e}_y + z {\bf e}_z)$,
and using the potential $\Phi$ and derivative multiplier $\Psi$ we have
the identity
\begin{equation}
\beta = \frac{\left<\Phi' u\right>
\left<\Psi u v + \frac{1}{Re} \Psi' u \right>}
{\left<\Phi \Psi\right>}.
\label{beta}
\end{equation}

The upper bound $\beta_b$ on the dissipation factor is obtained by first 
maximizing the right hand side of (\ref{beta}) over all normalized, 
divergence-free vector fields satisfying the boundary conditions, and then 
minimizing over all multiplier functions $\Psi \in H^1[0,1]$ satisfying 
homogeneous Dirichlet boundary conditions. 
Thus for any solution of the Navier-Stokes equations, $\beta \le \beta_b$ 
where the variational bound $\beta_b$ is the solution of the mini-max 
problem
\begin{equation}
\beta_b(Re) = \min_{\Psi} \max_{{\bf u}}
\frac{\left<\Phi' u\right>
\left<\Psi u v + \frac{1}{Re} \Psi' u \right>}
{\left<\Phi \Psi\right>}.
\label{minimax}
\end{equation}  
Note that while we explicitly display the Reynolds number dependence of 
$\beta_b$, it also depends the shape of the applied force---but {\it not} 
independently on the forcing amplitude $F$; the ratios in (\ref{beta1}), 
({\ref{beta}) and ({\ref{minimax}) are homogeneous in both $\phi$ and 
$\Phi$.

\subsection{Evaluting bounds}
From (\ref{minimax}) it follows immediately that $\beta_b(Re)$ is bounded 
by a function of the form $c_1 + c_2/Re$ for all Reynolds numbers, the 
analog of the result in \cite{Doering02}.
To see this, choose any convenient smooth multiplier function $\Psi$ 
(e.g., $\Phi$ and observe that elementary Cauchy-Schwarz and H\"older 
estimates (recalling the unit normalization of ${\bf u}$) give
\begin{equation}   
\left<\Phi' u\right> \le \left<\phi^2\right>^{1/2}, \ \ \
\left<\Psi u v \right> \le \frac{1}{2} \sup_{y\in[0,1]} \left| \Psi(y)
\right|, \ \ \
\left<\Psi' u \right> \le \left<\Psi'^2\right>^{1/2}
\end{equation}
so that
\begin{equation}
\beta_b(Re) = 
\frac{\left<\phi^2\right>^{1/2}\sup_{y\in[0,1]} \left| \Psi(y) \right|}
{2\left<\Phi \Psi\right>}
+ \frac{\left<\phi^2\right>^{1/2}\left<\Psi'^2\right>^{1/2}}
{\left<\Phi \Psi\right>} Re^{-1}.
\end{equation}
This simple analysis produces explicit expressions for the coefficients
$c_1$ and $c_2$ in a bound of the form $\beta \le c_1/Re + c_2$, 
displaying their functional dependence on the shape of the driving force.
In the following we will quantitatively and qualitatively improve this
upper bound, computing the {\it exact} solution of the infinite $Re$
limit of the mini-max problem.

To further estimate and evaluate $\beta_b(Re)$, note first that the 
boundary conditions together with incompressibility imply that
the $y$-component satisfies ${\overline v}(y) \equiv 0$ where the overbar
means horizontal and time average.
We decompose the $x$-component into a horizontal mean flow ${\overline 
u}(y)$ and a fluctuating remainder ${\tilde u} = u-{\overline u}$. 
Then the terms in the numerator of the ratio for $\beta_b$ reduce:
\begin{equation}
\left< \phi u \right> = \left< \phi {\overline u} \right>, \ \ \ \
\left< \Psi uv \right> = \left< \Psi {\tilde u} v \right>, \ \ \ \
\left< \Psi' u \right> = \left< \Psi' {\overline u} \right>.
\label{reduced}
\end{equation}
Let $\xi^2 = \langle{\overline u}^2\rangle$.
The normalization for the velocity field is
\begin{equation}
1 = \left<{\overline u}^2+{\tilde u}^2+v^2+w^2\right>
\ge \xi^2 + \left<{\tilde u}^2+v^2\right>,
\label{norm2} 
\end{equation}
so the terms in (\ref{reduced}) may be estimated
\begin{equation}
\left| \left< \phi u \right> \right| \le
\left< \phi^2 \right>^{1/2} \xi, \ \ \ \
\left| \left< \Psi uv \right> \right| \le
\frac{1}{2} \sup_{y\in[0,1]}\left|\Psi(y)\right| (1-\xi^2), \ \ \ \
\left| \left< \Psi' u \right> \right| \le
\left< \Psi'^2 \right>^{1/2} \xi.
\end{equation}
Hence for any choice of $\Psi$,
\begin{equation}
 \max_{{\bf u}} \frac{\left<\Phi' u\right> \left<\Psi u v + \frac{1}{Re}
\Psi' u \right>} {\left<\Phi \Psi\right>} \le
\max_{0\le\xi\le1} \frac{\left< \phi^2 \right>^{1/2}}{\left< \Phi \Psi
\right>} \xi \left[ \frac{1}{2} \sup_{y\in[0,1]} \left|\Psi(y)\right|
(1-\xi^2) + \frac{1}{Re} \left< \Psi'^2 \right>^{1/2} \xi \right].
\label{Sue}
\end{equation}
It is easy to find $\xi_m$, the maximizing value of $\xi$.
It is the solution of a quadratic equation in the interval $[0,1]$ for 
sufficiently high values of $Re$, or else it is $\xi_m = 1$ if 
\begin{equation}
Re \le 2 \frac{\left< \Psi'^2 \right>^{1/2}}
{\sup_{y\in[0,1]}\left|\Psi(y)\right|}.
\label{lowRe}
\end{equation}
When $\xi_m = 1$, the maximizing velocity field is a steady plane parallel 
flow, namely the Stokes flow for the given applied force.
The right hand side of (\ref{lowRe}) is $\ge 4$, providing, by a somewhat 
round-about derivation, a lower bound for the smallest possible critical 
Reynolds number of absolute stability of the steady plane parallel flow 
that is {\it uniform} in the shape of the applied shearing force.
For the purposes of the discussion here, however, we use the estimates 
above to bound $\beta_b$ as
\begin{eqnarray}
\beta_b &\le& \min_{\Psi} 
\frac{\left< \phi^2 \right>^{1/2}}{\left<\Phi \Psi \right>}  
\left[ \max_{0\le\xi\le1} \xi (1-\xi^2) \frac{1}{2}
\sup_{y\in[0,1]} \left|\Psi(y)\right| + \max_{0\le\xi\le1} \xi^2
\frac{1}{Re} \left< \Psi'^2 \right>^{1/2} \right] \nonumber \\
&=& \min_{\Psi} \frac{\left< \phi^2 \right>^{1/2}}{\left<\Phi \Psi
\right>} \left[ \frac{1}{\sqrt{27}} \sup_{y\in[0,1]} \left|\Psi(y)\right|
+ \frac{1}{Re} \left< \Psi'^2 \right>^{1/2} \right].
\label{imp}
\end{eqnarray}
This leads to improved estimates, in terms of variational problems for an 
optimal multiplier $\Psi$ and for $c_2$ in a bound of the form $\beta_b 
\le c_1/Re + c_2$.
As we will now show, the variational expression above for $c_2$ is sharp 
at high Reynolds numbers.
That is, the $Re \rightarrow \infty$ limit of the extremization problem 
for the optimal $\Psi$ and $\beta_b$ can be solved exactly.

Define
\begin{equation}
\beta_b(\infty) = \min_{\Psi} \max_{{\bf u}} 
\frac{{\left<\Phi' u \right>}{\left<\Psi uv \right>}}
{{\left<\Phi \Psi \right>}}.
\end{equation}
First we will evaluate $\beta_b(\infty)$, and then we will prove that 
\begin{equation}
\limsup_{Re \rightarrow \infty} \beta_b(Re) \le \beta_b(\infty).
\end{equation}
We accomplish this through a series of two lemmas and a theorem.

\bigskip

\noindent
{\bf Lemma 1:} If $\phi = -\Phi' \in L^2[0,1]$ and $\Psi \in H^1[0,1]$ 
satisfies $\Psi(0)=0=\Psi(1)$, then
\begin{equation}
\max_{{\bf u}} \left<\Phi' u \right> \left<\Psi uv \right> =
\frac{1}{\sqrt{27}} \left< \phi^2 \right>^{1/2} \sup_{y\in[0,1]}
\left|\Psi(y)\right|
\end{equation}
where the velocity fields ${\bf u}$ are divergence-free and unit
normalized in $L^2$, periodic in downstream ($x$) and spanwise ($z$)
directions and free-slip in the normal ($y$) direction.

\smallskip

\noindent
{\it Proof:} We may take $\Psi \ne 0$.
The calculation for (\ref{imp}) already established that the proposed 
answer is an upper bound to this variational problem, so all we must do is 
display a sequence of acceptable test fields ${\bf u}$ that approach the 
bound. 
Note that any nonvanishing $\Psi \in H^1[0,1]$ satisfying the homogeneous 
Dirichlet conditions is uniformly continuous and its extremum is realized 
at a (not necessarily unique) point $y_m$ in the open interval $(0,1)$.
Consider the unit-normalized divergence-free vector field ${\bf u}_k$ with
components
\begin{eqnarray}
u_k &=& g_k(y) \sqrt{2} \sin{kz} + \frac{1}{\sqrt{3}}
\frac{\phi(y)}{\sqrt{\left< \phi^2 \right>}} \nonumber \\
v_k &=& \pm g_k(y) \sqrt{2} \sin{kz} \nonumber \\
w_k &=& \frac{1}{k} g_k'(y) \sqrt{2} \cos{kz}
\end{eqnarray}
where $g_k(y)^2$ is a smooth approximation of a $\delta$-function with
compact support centered on $y_m$ and normalized according to
\begin{equation}
\frac{1}{3} = \left< g_k^2 + \frac{g_k'^2}{2k^2} \right>.
\end{equation}
The wavenumber $k \ne 0$ is adjustable, and for each value of $k$ we have
$\left<\phi u_k \right> = \frac{1}{\sqrt{3}} \left<\phi^2 \right>^{1/2}$.
Now $\left<\Psi u_kv_k \right> = \left<\Psi \overline{u_kv_k}
\right>$ and we may concentrate $\overline{u_kv_k}(y) =
\pm g_k(y)^2$ as tightly as desired around $y_m$ by taking $k$ large, in
which case $\left<\Psi u_kv_k \right> \rightarrow \left| \Psi(y_m)
\right| \left< g_k^2 \right>.$
Moreover, $\left< g_k^2 \right> \rightarrow 1/3$ as 
$k \rightarrow \infty$.
Hence there exists a sequence of test fields ${\bf u}_k$ for which
$\left<\phi u_k \right> \left<\Psi u_kv_k \right>$ approaches the upper
bound.
{\it QED}.

\bigskip

\noindent
{\bf Lemma 2:} If $\Phi \in H^1[0,1]$ and 
$\mbox{sign}\left[\Phi(y)\right]$ has a finite number of discontinuities, 
then 
\begin{equation} 
\min_{\Psi \in H^1[0,1]} \frac{\sup_{y \in [0,1]} \left| \Psi(y) \right|}
{\left< \Phi \Psi \right>} = \frac{1}{\left< \left| \Phi \right| \right>}. 
\end{equation}
The minimizing function is $\Psi_m(y) = \mbox{sign}\left[\Phi(y)\right]$
which is not in $H^1[0,1]$, but it is the pointwise limit of a sequence of
functions in $H^1$.

\smallskip

\noindent
{\it Proof:} Note that
\begin{equation}
\left< \Phi \Psi \right> \le \sup_{y\in[0,1]} \left| \Psi(y) \right| 
\left< \left| \Phi \right| \right>
\end{equation}
so the proposed answer is a lower bound to the minimum, and the function
$\Psi_m(y) = \mbox{sign}\left[\Phi(y)\right]$ saturates this bound.
Then it is straightforward to mollify the finite number of discontinuties 
in $\Psi_m$ to produce a sequence of $H^1$ functions converging pointwise 
to $\Psi_m$.
Let the number of discontinuities of $\Psi_m(y)$ be $N$, located in order 
at $y_n$.
We may smooth $\Psi_m$ by introducing a finite slope near each $y_n$ to 
produce the regulated function $\Psi_{\delta}(y)$ sketched in 
figure~\ref{fig1}.
The mollified $\Psi_{\delta}(y)$ is linear with slope $\pm \delta^{-1}$ 
inside all the intervals $2\delta$ around each $y_n$ and within $\delta$ 
of the ends of the interval at $y=0$ and $1$.
Because $\Phi \in H^1[0,1]$, $\Phi = {\cal O}(\sqrt{|y-y_n|})$ near the
isolated zeros $y_n$ where $\Psi_m$ jumps.
Thus $\left< \Phi \Psi_{\delta} \right> = \left< \left| \Phi \right| \right> 
\left[1 - {\cal O}(\delta^{3/2})\right]$.
Although $\left<(\Psi'_{\delta})^2 \right> = {\cal O}(\delta^{-1})$, for 
each $y \in [0,1]$, $\Psi_{\delta}(y) \rightarrow \Psi_m(y)$ as $\delta
\rightarrow 0$.
{\it QED}.
\begin{figure}
\begin{center}
\epsfig{file=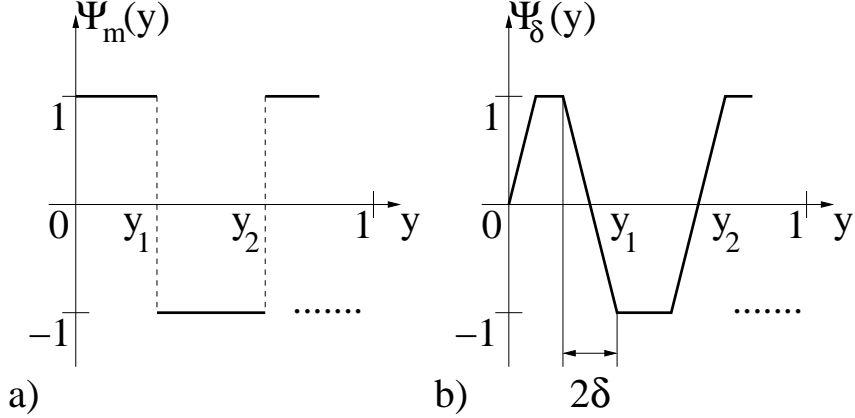,width=6cm,angle=-90}
\end{center}
\caption{Boundary layer regularization for the optimal multiplier 
$\Psi_m$.
(a) Sketch of $\Psi_m(y)$ with a finite number of jumps in [0,1] at $y = 
0, y_1, y_2, \dots, 1$.
(b) Sketch of the mollified $\Psi_{\delta}(y)$ that converges pointwise to 
$\Psi_m(y)$ for $\delta\to 0$.}
\label{fig1}
\end{figure}

\bigskip

\noindent
{\bf Theorem:} If $\Phi \in H^1[0,1]$ such that
$\mbox{sign}\left[\Phi(y)\right]$ has a finite number of jump 
discontinuities, then
\begin{equation}
\beta_b(\infty) \ = \ \frac{1}{\sqrt{27}}\
\frac{\sqrt{\left< \phi^2 \right>}}{\left< \left| \Phi \right| \right>}
\label{e328}
\end{equation}
and
\begin{equation}
\beta_b(Re) \ \le \ \beta_b(\infty) \ + \ {\cal O}(Re^{-3/4})
\label{e329}
\end{equation}
as $Re \rightarrow \infty$.

\smallskip

\noindent
{\it Proof:}
Lemmas 1 and 2 establish the value of $\beta_b(\infty)$ in (\ref{e328}).
To establish (\ref{e329}), recall from the proof of Lemma 2 that
$\left< \Phi \Psi_{\delta} \right> = \left< \left| \Phi \right| \right> 
\left[1 - {\cal O}(\delta^{3/2})\right]$ and
$\left<(\Psi'_{\delta})^2 \right> = {\cal O}(\delta^{-1})$.
Using these facts together with Lemma 1,
\begin{eqnarray}
\beta_b(Re) &=& \min_{\Psi} \max_{{\bf u}} \frac{\left<\Phi' u\right>
\left<\Psi u v + \frac{1}{Re} \Psi' u \right>} {\left<\Phi \Psi\right>}
\nonumber \\
&\le& \min_{\delta} \max_{{\bf u}} \frac{\left<\Phi' u\right>
\left<\Psi_{\delta} u v + \frac{1}{Re} \Psi'_{\delta} u \right>}
{\left<\Phi \Psi_{\delta}\right>} 
\nonumber \\
&\le& \min_{\delta} \max_{{\bf u}} \frac{\left<\Phi' u\right>
\left(\left<\Psi_{\delta} u v\right> + \frac{1}{Re} {\cal O} 
(\delta^{-1/2}) \right)}{\left<\Phi \Psi_{\delta}\right>}
\nonumber \\
&\le& \min_{\delta} \left(\frac{1}{\sqrt{27}}
\frac{\left<\phi^2\right>^{1/2}}{\left< \left| \Phi \right| \right>}
\left[1+{\cal O}(\delta^{3/2})\right]\left[1+\frac{1}{Re} 
{\cal O}(\delta^{-1/2})\right] \right).
\end{eqnarray}
Choosing $\delta = {\cal O}(Re^{-1/2})$ establishes the result.
{\it QED}.  

\bigskip

We make three short technical remarks here:

\smallskip

\noindent
(i) Although we only showed that $\limsup_{Re \rightarrow \infty} 
\beta_b(Re) \le \beta_b(\infty)$, it is natural to conjecture that at 
finite $Re$ the optimal multiplier $\Psi$ actually looks like the 
mollified multipliers $\Psi_{\delta}$ and that $\lim_{Re \rightarrow 
\infty} \beta_b(Re) = \beta_b(\infty)$.
But this remains to be proven.

\smallskip

\noindent
(ii) The ${\cal O}(Re^{-3/4})$ rate of approach to the $Re \rightarrow 
\infty$ limit in the theorem is not optimal for smoother shape functions.  
This is easy to see by repeating the proof of the theorem assuming, say, 
that $\phi \in H^1$ so that $\Phi$ has a bounded derivative and behaves 
linearly (rather than as a square root) near its zeros.  
That generic linear behavior leads to a faster ${\cal O}(Re^{-4/5})$ rate.

\smallskip

\noindent
(iii) The hypothesis of a finite number of zeros in $\Phi$ is probably not 
really necessary given $\Phi \in H^1$; we invoke it here for simplicity of 
the proofs only.
In any case, for the applications we have in mind, $\phi$ and $\Phi$ will 
actually be extremely smooth (composed, for example, of a finite number of 
Fourier components) so the theorem as stated and proved here serves our 
purposes.

\section{Comparison with numerical results and discussion}

Direct numerical simulations (DNS) in this geometry with these kinds of 
forces are possible in Fourier space thanks to the free-slip boundary 
conditions. 
For computations we used the pseudospectral code developed in 
\cite{Schumacher2000} and \cite{Schumacher2001} with numerical resolution 
of $256\times 65\times 256$ grid points.
The steady volume forcing density, ${\bf f}({\bf x})$, was chosen such 
that a laminar (and linearly stable!) shear flow profile ${\bf u}_0({\bf 
x})=- U_0\cos(\pi y/\ell) {\bf e}_x$ could be sustained.
From the Navier-Stokes equations (\ref{nseq}) it follows for this 
plane-parallel shear flow that
\begin{eqnarray}
{\bf f}({\bf x})= -\frac{\nu U_0 \pi^2}{\ell^2} \cos(\pi y/\ell) {\bf e}_x 
= F \phi\left(\frac{y}{\ell}\right){\bf e}_x
\end{eqnarray}
with shape $\phi(\eta)=\sqrt{2} \cos{\pi \eta}$ and amplitude $F=-\nu 
U_0\pi^2/\sqrt{2}\ell^2$.
The flow can be considered a Kolmogorov flow (see \cite{Borue96} and 
\cite{Childress2001}) with additional symmetry constraints in the normal 
($y$) direction. 
The aspect ratio and scales for the calculations were 
$L_x/\ell=L_z/\ell=2\pi$.
The Grashof number for absolute (energy) stability of the steady 
plane-parallel flow with this force shape is $Gr_c = 68$ where the 
Reynolds number is less than 7; the simulations were carried out well 
above this value, for $Gr$ varying between $4900$ and $59200$.

Mathematical results for the shape function $\phi(y) = -\sqrt{2}\cos{\pi 
y}$ (equivalently $\Phi(y) = \pi^{-1} \sqrt{2} \sin{\pi y}$) are shown in 
Figure 2 along with the DNS results. 
In contrast to shear flows driven by rigid walls where the friction 
(dissipation) factor tends to decrease with increasing Reynolds number, 
here we observe a slight increase. 
The numerical values for $\beta$ are about a factor 3 below the upper 
bound.
This is a significantly better comparison of the data and the bounds than 
for turbulent Couette flow where the discrepancy is a factor 10 at $Re 
\approx 10^6$. 

The mean profiles of the streamwise velocity component for two different 
Reynolds numbers are shown in the inset in Figure 2. 
It is interesting to note that even though the force shape is nonlinear 
across the layer, the mean profile is relatively linear with mean shear 
nearly constant outside boundary layers near the no-slip walls.
The high Reynolds number limit of the optimal multiplier $\psi_m(y)$ is
piecewise linear with constant magnitude of its slope function;
$|\psi_m'(y)| = |\Psi_m(y)| = 1$ away from the corners.
We point out the similarity here with the observed mean profiles for the
single example we have at hand.


\begin{figure}
\begin{center}
\epsfig{file=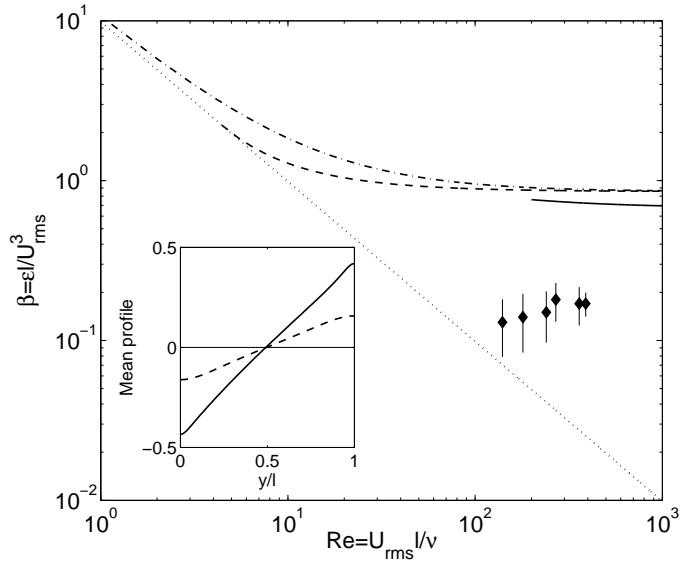,width=9cm}
\end{center}
\caption{The dissipation factor $\beta=\epsilon \ell/U^3$ as a function of 
the Reynolds number $Re=U \ell/\nu$ for the force shape function
$\phi(y/\ell) = -\sqrt{2} \cos{\pi y/\ell}$.
The results of the direct numerical simulations are indicated by diamonds 
with error bars due to standard deviation $\pm\sigma_{\beta}$ where 
$\sigma_{\beta}=\beta(\sigma_{\epsilon}/\epsilon + 
3\sigma_{U_{rms}}/U_{rms})$.
The lower dotted line is the dissipation in the steady laminar flow which 
is a lower limit to the dissipation factor for any (statistically) 
stationary flow with this force.
The three upper bounds, from top to bottom, are the estimate in 
(\ref{imp}) with the choice $\Psi = \Phi$ (dash-dot line), the estimate in 
(\ref{Sue}) with the exact maximization over $\xi$ followed by the choice 
$\Psi = \Phi$ (dashed line), and the optimal value $\beta_b(\infty) = 
\pi^2/\sqrt{216}$ from the theorem with the rigorous ${\cal O}(Re^{-4/5})$ 
approach added on (solid).
The optimal bound for the infinite $Re$ limit is a $22\%$ improvement 
below the infinite $Re$ limit of the bound with $\Psi = \Phi$.
The mean flow profiles $\overline{u_x}(y)$ for the simulations with the 
smallest (dashed) and largest (solid) $Re$ are shown in the inset.}
\label{fig2}
\end{figure}


It will be very interesting to study the bounds $\beta_b(Re)$ as well as 
the optimal multiplier functions at finite Reynolds numbers for a variety 
of force shape functions $\phi$.
This is because while the behavior of the bound on $\beta$ is similar in 
structure to the observed experimental and computational values 
(\cite{Sreeni84,Sreeni98}), it remains an open question how the high-$Re$ 
value of $\beta$ depends on the details of the driving.   
There are some features we can anticipate right away, though.
Assuming that the structure of $\Psi_m(y)$ persists for large but finite 
Reynolds numbers, the high-$Re$ optimal multiplier is a simple but 
interesting nonlinear functional of the shape function of the driving 
force.
While the plane-parallel Stokes flow profile $U_{Stokes}(y)$ is a linear 
functional of the shape function,
\begin{equation}
U_{Stokes}(y) \sim \int_0^y{dy' \left[ \int_0^{y'}{dy'' \phi(y'')} \right] 
} + C,
\end{equation}
the (infinite $Re$) optimal multiplier comes from a curiously 
similar---but highly nonlinear---formula:
\begin{equation}
\psi_m(y) \sim \int_0^y{dy' \mbox{sign} \left[ \int_0^{y'}{dy''
\phi(y'')} \right] } + C.
\end{equation}
This expression for $\psi_m$ displays bifurcations as a functional of the 
shape function $\phi$.
That is, for some shape functions, variations in $\phi$ may result in no 
change at all in the associated high-$Re$ optimal multiplier, while at    
other configurations small changes in $\phi$ can produce large changes in 
$\psi_m$ (such as the number of ``kinks'' in the multiplier profile).
Whether or not this kind of effect reflects any features of high Reynolds 
number mean profiles for shear turbulence driven by other shaped forces 
remains to be seen.

To summarize, in this paper we have derived and analyzed a variational 
mini-max problem for upper bounds on the energy dissipation rate valid for 
both low and high Reynolds number (including turbulent) body-forced shear 
flows.
We find that the maximizing flow fields are characterized by streamwise 
vorticies concentrated near the maximal shear in an auxillary 
``multiplier" profile, analogous to the ``background" profile utilized in 
\cite{Doering94}, \cite{Nicodemus98}, \cite{Kerswell98} and 
\cite{Childress2001}.
We solved the optimal high $Re$ mini-max problem exactly and compared the 
results with data from direct numerical simulations for a specific choice 
of forcing.
We observed that the high $Re$ bound is only about a factor of three above 
the data, and also that the high $Re$ optimal multipier shares some  
qualitative features with the measured mean flow profiles.
Future work in this area will include investigations for other force 
shapes, as well as the improvement of rigorous bounds by exact numerical
evaluation and/or by the inclusion of a balance parameter 
(see \cite{Nicodemus97}) in the variational problem. 
Finally, we remark that although we have carried out the bounding analysis 
with a steady driving for the flow, there is no obstruction to the 
inclusion of time-dependent forcing in the model.

\begin{acknowledgments}
The numerical simulations were done on a Cray SV1ex at the John von 
Neumann-Institut f\"ur Computing at the Forschungszentrum J\"ulich; we
are grateful for that support.
This research was supported by the Deutsche Forschungsgemeinschaft
and the US-NSF.  
Some of this work was completed while one of us (CRD) was resident at the
Geophysical Fluid Dynamics Program at Woods Hole Oceanographic Institution 
during the summer of 2002, and helpful and stimulating discussions with N. 
Balmforth, P. Constantin, J. Keller and R. Kerswell are gratefully 
acknowledged.
\end{acknowledgments}

\end{document}